\documentclass[aps,twocolumn,floats,prd,nofootinbib,superscriptaddress,10pt]{revtex4-1}
\usepackage[utf8]{inputenc}
\usepackage{bm}
\usepackage{amsmath}
\usepackage{comment}
\usepackage{color}
\def\blue{\textcolor{black}}
\usepackage{graphicx,amsmath,amsfonts,amssymb
}

\newcommand{\changeJ}[1]{{\textcolor{black}{#1}}}

\begin{document}

\title{Shedding light on the small-scale crisis with CMB
spectral distortions}

\author{Tomohiro Nakama}

\affiliation{Department of Physics and Astronomy, Johns Hopkins
     University, 3400 N.\ Charles St., Baltimore, MD 21218, USA}

\author{Jens Chluba}

\affiliation{Jodrell Bank Centre for Astrophysics, School of
Physics and Astronomy, University of Manchester, Oxford Road,
Manchester M13 9PL, UK}

\author{Marc Kamionkowski}

\affiliation{Department of Physics and Astronomy, Johns Hopkins
     University, 3400 N.\ Charles St., Baltimore, MD 21218, USA}

\begin{abstract}
The small-scale crisis, discrepancies between observations and
N-body simulations, may imply suppressed matter
fluctuations on subgalactic distance scales.  Such a suppression
could be caused by some early-\blue{universe} mechanism (e.g., broken
scale invariance during inflation), leading to
a modification of the primordial power spectrum at the onset of
the radiation-domination era.  Alternatively,
it may be due to nontrivial dark-matter properties (e.g., new
dark-matter interactions or warm dark matter) that affect the \blue{matter}
power spectrum at late times, during radiation domination, after
the perturbations re-enter
the horizon. We show that early- and late-time suppression
mechanisms can be distinguished by measurement of the $\mu$
distortion to the frequency spectrum of the cosmic microwave
background. This is because the $\mu$ distortion is
suppressed, if the power suppression is primordial, relative to
the value expected from the dissipation of standard
nearly scale-invariant fluctuations.  We emphasize that the
standard prediction of the $\mu$ distortion remains unchanged
in late-time scenarios even if the dark-matter effects occur
before or during the era (redshifts $5\times 10^4 \lesssim z
\lesssim 2\times 10^6$) at which $\mu$ distortions are
generated.
\end{abstract}

\maketitle

\section{Introduction}

The canonical $\Lambda$CDM model of cosmic structure formation,
in which structures grow from a nearly scale-invariant spectrum
of primordial adiabatic perturbations, has achieved many
successes.  Still, there are discrepancies between observations
on subgalactic scales and predictions from N-body simulations of
structure and galaxy formation.  We refer to these discrepancies
collectively as the ``small-scale crisis'' (see
Ref.~\cite{Weinberg:2013aya} for a review), which includes the
missing-satellite problem \cite{Moore:1999nt}, the cusp-vs-core
problem \cite{Moore:1999gc}, and the too-big-to-fail problem
\cite{BoylanKolchin:2011de}.  While further elucidation of the
relevant baryonic physics may help solve these problems
\cite{Bullock:2000wn,Benson:2001at,Simon:2007dq},
the small-scale crisis may also imply a suppression of density
fluctuations below or around subgalactic scales
\cite{Weinberg:2013aya}.

Exotic mechanisms to suppress small-scale power can be
classified into those that involve
modification of the primordial power in the early Universe and
those that suppress power at later times.  Examples of
\textit{primordial suppression} are discussed in
Refs.~\cite{Kamionkowski:1999vp,Yokoyama:2000tz,Zentner:2002xt,Ashoorioon:2006wc,Kobayashi:2010pz},
and predict that subgalactic primordial \blue{adiabatic} perturbations are
suppressed at the {\it onset} of the radiation-dominated 
era.  Examples of {\it late-time} suppression are those
discussed, e.g., in Refs.~\cite{Kawasaki:1996wc,Lin:2000qq, Hisano:2000dz,
Sigurdson:2003vy, Profumo:2004qt, Cembranos:2005us,
Kaplinghat:2005sy,
Kusenko:2009up, Kohri:2009mi, Aarssen:2012fx,Kamada:2013sh,
Binder:2016pnr,Bringmann:2016ilk,Hu:2000ke}; in such
scenarios, subgalactic \blue{matter} perturbations are present at the
onset of the radiation-dominated era but then suppressed at
later times, \blue{after} the relevant scales re-enter the horizon.
These mechanisms include free streaming of dark-matter \changeJ{(DM)} particles
(e.g., as in warm dark matter, WDM) or interactions between \changeJ{DM}
and standard-model or hidden particles.

There has been exploration of different astrophysical
consequences of suppressed small-scale power, and constraints to
models are already being derived.  For example, \blue{WDM}
provides a particularly well-studied example of
a late-time-suppression mechanism, and some recent work
\cite{Viel:2013apy,Schneider:2013wwa} suggests tensions between
WDM solutions to the small-scale crisis and observations.
Even if a WDM-only scenario for resolving the small-scale
crisis is in tension with Lyman-$\alpha$
observations, a mixture of cold DM and WDM (mixed dark matter)
can still evade Lyman-$\alpha$ constraints while helping to
solve problems associated with the small-scale crisis
\cite{Boyarsky:2008xj,Anderhalden:2012qt,Harada:2014lma,Kamada:2016vsc}.
There is thus still good reason to consider solutions to
the small-scale crisis \blue{based on \changeJ{power} suppression}.

For a primordial suppression, the shape of the suppressed
spectrum on small scales depends on unknown and largely
unconstrained early-\blue{universe} physics.  In the framework of
single-field inflation, for instance, it is determined by the
slope of the inflaton potential
\cite{Kamionkowski:1999vp}. Hence, even though a wide variety of
late-time suppression mechanisms that predict different
shapes for a suppressed power spectrum have been proposed,
measurement of the {\it shape} of the \blue{late-time matter power spectrum on small scales}
cannot really distinguish whether the suppression is primordial
or late time.

Here we point out that primordial and late-time
suppression mechanisms can be distinguished by the cosmic
microwave background (CMB) $\mu$ distortion
\cite{Zeldovich:1969ff,
Burigana:1991,Hu:1992dc,Chluba:2011,Hu:1994, Chluba:2012gq,
Dent:2012ne, Chluba:2012we, Khatri:2013dha, 
Tashiro:2014pga, Khatri:2015tla
}.
Such $\mu$ distortions are produced by heating of the primordial
plasma from dissipation of small-wavelength fluctuations.  The
Fourier modes that contribute to the $\mu$ distortions have
comoving wavenumbers $50\,{\rm Mpc} \lesssim k \lesssim
10^4\,{\rm Mpc}$, and the $\mu$ distortion arises when \blue{these}
dissipate at redshifts $5\times 10^4 \lesssim z \lesssim 2\times
10^6$ \blue{(the $\mu$ era)}.    In the standard scenario, where the
nearly scale-invariant spectrum of perturbations seen in
the \blue{CMB} \cite{Ade:2015xua} is
extrapolated to smaller scales (as
motivated by Occam's razor and the simplest models of
inflation), the induced $\mu$ distortion is $\mu\simeq 2\times
10^{-8}$ \cite{Chluba:2012gq,
Chluba:2013pya,Cabass:2016giw,Chluba:2016bvg}.

For primordial suppression, the value of $\mu$ will be reduced
relative to that expected from the standard almost-scale-invariant
spectrum.  If the suppression is strong enough, the $\mu$
parameter could even take a \textit{negative} value,
$\mu_{\mathrm{BE}}\simeq -3\times 10^{-9}$
\cite{Chluba:2011,Pajer:2013oca,Khatri:2011aj},
due to the continuous extraction of energy from \blue{CMB} photons by the
nonrelativistic baryons to which they are coupled.\blue{\footnote{This arises
because the baryons \blue{alone would cool} as $T_b \propto a^{-2}$
with the scale factor $a$, as opposed to \changeJ{$T_\gamma \propto
a^{-1}$ for photons}.
}
}

In contrast, for late-time suppression of small-scale
perturbations, there are still primordial perturbations to be
dissipated, and so the standard prediction is unmodified.  The
argument is not entirely trivial, as in many late-time
scenarios, the suppression of the matter power spectrum occurs
during the $\mu$ era.  For example, in the
charged-particle-decay scenario \cite{Sigurdson:2003vy},
suppression of the matter power spectrum occurs until roughly
3.5 years after the Big Bang, at redshifts $z\simeq 5\times10^5$, right
when the $\mu$ distortion is being produced.  This timescale is
actually fairly generic, as this is the redshift at which
subgalactic scales are entering the horizon.  The crucial point
is that the matter density is negligible compared with the radiation 
density during the $\mu$ era.  There can thus be dramatic smoothing of the matter
distribution with little effect on the radiation-density
perturbations. Similar arguments apply to the $y$ distortion, \changeJ{which is created later at $z\lesssim 5\times 10^4$ \cite{Zeldovich:1969ff, Burigana:1991}. Here in addition, distortions are sourced by bulk flows at second order in the baryon velocity, $v$. However, these contributions are subdominant \cite{Chluba:2012gq} relative to larger energy release from first stars and structure formation, as also} recently discussed in Ref.~\cite{Sarkar:2017vls}.  \changeJ{For the same
reasons} the effects of primordial dark-matter isocurvature perturbations on spectral distortions are limited \cite{Chluba:2013dna}.
 
In the next section, we calculate \blue{the value for} $\mu$ assuming a step-type
suppression of primordial power below subgalactic scales, and we
conclude in Sec. III.

\section{PRIMORDIAL SUPPRESSION AND CMB $\mu$ DISTORTION}
\begin{figure}[htbp]
\begin{center}
\includegraphics[width=8.5cm,keepaspectratio,clip]{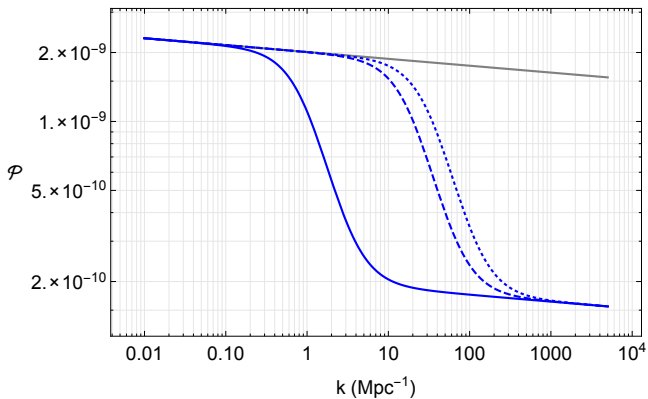}
\end{center}
\caption{Examples of primordial power spectra suppressed below
     subgalactic scales [\,Eq.\!\! (\ref{power})\,] considered
     in this paper.  \blue{For the blue curves, $\alpha=1$, and from bottom to top \changeJ{we have $k_s=\{1,20,35\}\mathrm{Mpc}^{-1}$}. The gray curve corresponds to the \changeJ{standard spectrum} ${\cal P}_{\mathrm{st}}$ of Eq.\ (\protect\ref{power}) ($\alpha=0$). }
}
\label{power3}
\end{figure}

We relate the primordial power suppression to the $\mu$
distortion as follows. We employ the following description of
primordial suppression in terms of the dimensionless primordial
curvature power spectrum:
\begin{eqnarray}
{\cal P}(k)&=&{\cal P}_{\mathrm{st}}(k)\tilde{{\cal P}}(k),\quad
{\cal P}_{\mathrm{st}}(k)={\cal
A}\left(\frac{k}{k_p}\right)^{n_s-1}, \nonumber \\
\tilde{{\cal
P}}(k)&=&\frac{1+10^{-\alpha}}{2}-\frac{1-10^{-\alpha}}{2}\tanh\left(\log\frac{k}{k_s}\right).
\label{power}
\end{eqnarray}
That is, the power is suppressed by $10^{-\alpha}$ for $k\gtrsim
k_s$, relative to the standard spectrum ${\cal P}_{\mathrm{st}}$
with parameters ${\cal A}=2.2\times 10^{-9},
\,k_p=0.05\,\mathrm{Mpc}^{-1}$ and $n_s=0.97$
\cite{Ade:2015xua}.
Examples of the suppressed primordial spectra are shown in
Fig.\ \ref{power3}, where we take
$k_s=1,\,20$ and
$35\,\mathrm{Mpc}^{-1}$, relevant for small-scale
problems. The above step-type suppression would lead to a
step-type suppression of the matter spectrum at low redshifts,
similarly to mixed-DM scenarios
\cite{Boyarsky:2008xj,Anderhalden:2012qt}. Hence, we can refer
to those studies to gain insight into structure
formation for the suppressed primordial spectrum we consider here. However,
establishing a precise link between our spectrum and
different aspects of the small-scale crisis is beyond the scope
of this work, \blue{at least} because of potentially important
baryonic processes. Thus, we treat $\alpha$ and $k_s$ as free
parameters and only illustrate that $\mu$ can be significantly
smaller than the expected standard value, $\mu\simeq 2\times
10^{-8}$. 

Additional information about the precise
position of the transition scale might be accessible with
future measurements of the exact spectral-distortion shape
\cite{Chluba:2012gq, Chluba:2012we,Chluba:2013pya}.
However, even if we were only to observe a significant
suppression of $\mu$, without additional information \blue{about} the
spectral-distortion shape, we could connect the small-scale
crisis to primordial suppression.  Ultimately, it will be
instructive to investigate small-scale problems with
simulation of structure formation with the various primordial
spectra which are consistent with, e.g., simultaneous
constraints from the Lyman-$\alpha$ forest and $\mu$ (and
possibly taking into account baryonic processes). 

The $\mu$ distortion can be estimated \cite{Chluba:2015bqa}\footnote{Here we assume no other
energy injection mechanisms in the early Universe exist, such as
evaporating primordial black holes, decaying/annihilating
particles, cosmic strings, primordial magnetic fields and
axionlike particles (see Ref.\ \protect\cite{Tashiro:2014pga}
for an overview).} as
$\mu = \mu_{\mathrm{ac}}+\mu_{\mathrm{BE}}$ with
$\mu_{\mathrm{BE}} \simeq -3\times 10^{-9}$ and
\begin{equation}
\mu_{\mathrm{ac}}\simeq \int_{k_{\mathrm{min}}}^\infty
\frac{dk}{k}{\cal P}(k)W_{\mu}(k),
\end{equation}
with
\begin{eqnarray}
W_\mu(k)&\simeq& 2.8A^2
\left[
\exp\left(
-\frac{[\hat{k}/1360]^2}{1+[\hat{k}/260]^{0.3}+\hat{k}/340}
\right) \right. \nonumber \\
 & & \left. -\exp
\left(
-\left[\frac{\hat{k}}{32}\right]^2\right)
\right],
\end{eqnarray}
where $k_{\mathrm{min}}\simeq 1\mathrm{Mpc}^{-1}$, $A\simeq 0.9$
and $\hat{k}=k \,\mathrm{Mpc}$.
This approximation is accurate at the $\simeq 20\%$ level and
slightly underestimates the recovered value for
$\mu$ \cite{Chluba:2016bvg}. Hence, we renormalize the
above window function $W_\mu$ so that $\mu_{\mathrm{ac}}\simeq
2.3\times 10^{-8}$ when $\alpha=0$ (i.e. standard fluctuations),
which is sufficient for our purposes. 

\begin{figure}[htbp]
\begin{center}
\includegraphics[width=8.5cm,keepaspectratio,clip]{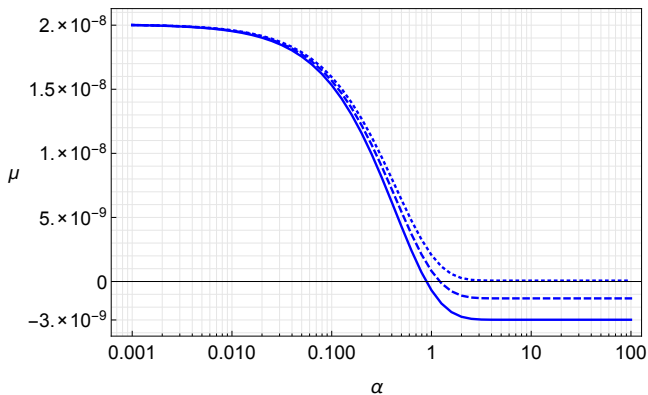}
\end{center}
\caption{The dependence of $\mu$ distortions on $\alpha$, which
     controls a step-type primordial suppression [see
     Eq.\! (\protect\ref{power})]. From bottom to top, the
     suppression wave number is $k_s=\{1,20,35\}\mathrm{Mpc}^{-1}$.  As $\alpha\rightarrow 0$,
     $\mu$ approaches $\simeq 2\times 10^{-8}$, the value
     mostly determined by the dissipation of the standard
     almost-scale-invariant fluctuations. In contrast, if $k_s$
     is relevant to the small-scale crisis and if $\alpha$ is
     sufficiently large, $\mu$ can be negative, approaching
     $\mu_{\mathrm{BE}}\simeq -3\times 10^{-9}$ for $k_s\sim
     1\,\mathrm{Mpc}^{-1}$, determined by the energy extraction
     from photons to baryons due to their coupling.}
\label{main}
\end{figure}

The values of $\mu$ as a function of $\alpha$ are shown in
Fig.\ \ref{main}. When $k_s$ is close to $\sim
1\,\mathrm{Mpc}^{-1}$ and  $\alpha$ is sufficiently large, $\mu$
becomes negative, approaching $\mu_{\mathrm{BE}}$.
For $k_s\simeq 35\,\mathrm{Mpc}^{-1}$, the asymptotic
value is $\mu\simeq 0$; that is, the energy injection due to the
dissipation of sound waves and energy extraction due to
interactions between photons and baryons are \changeJ{roughly} balanced. 
If in the future $\mu$ is constrained to be smaller than what is
expected ($\mu\simeq \mu_{\mathrm{ac}}\simeq 2\times 10^{-8}$),
from the dissipation of the standard fluctuations
(in the figure this corresponds to the limit
$\alpha\rightarrow 0$), then it could serve as a smoking gun for
some primordial suppression thereby possibly explaining
the small-scale crisis.  

\section{CONCLUSION}

\vspace{-3mm}
The small-scale crisis of $\Lambda$CDM may imply suppressed
matter fluctuations on subgalactic scales. Such a suppression
could result from some new physics that operates during
inflation or could be the consequence of new dark-matter physics
that operates at later times, \blue{after} the relevant distance scales
re-enter the horizon during radiation domination.
Although the primordial and late-time suppression
mechanisms are expected to impact structure formation in a
similar fashion, we show here that they could be in principle
distinguished by measurement of the $\mu$ distortion to the CMB
frequency spectrum. This is
because $\mu$ may be significantly reduced relative to the
canonical value $\mu\simeq2\times 10^{-8}$ if subgalactic power
suppression is primordial.  For power suppression sufficiently
significant, $\mu$ could even become negative as a consequence of the
transfer of energy from photons to baryons.  On the other hand, for a
late-time suppression, the CMB $\mu$ distortion would not be
affected notably since it is mostly determined by primordial
fluctuations rather than subhorizon dynamics of DM fluctuations
during the radiation-dominated era. Thus, for a late-time
suppression, $\mu$ is not expected to differ significantly from
the standard positive value.

If $\mu$ is found to be unexpectedly small or negative by future
high-sensitivity experiments measuring the energy spectrum of
CMB photons, it may serve as a \textit{smoking gun} for a
primordial suppression.  \blue{Note also that the negative contribution to $\mu$ can, in principle, be even smaller than $\mu_{\mathrm{BE}}$
due to direct or indirect thermal coupling of non-relativistic
DM with photons, since in this case more energy is extracted
from photons to DM to maintain thermal equilibrium
\protect\cite{Ali-Haimoud:2015pwa}.}
If \blue{on the other hand} the standard prediction for $\mu$ is
verified, then it suggests that the small-scale crisis has to do
with late-time physics.  If we find $\mu$ to have the
standard value, then another possibility, which we leave
for future work, is that a matter-radiation isocurvature
perturbation, correlated with the adiabatic perturbation,
suppressed matter perturbations on small scales while preserving
the primordial curvature (and thus radiation) perturbation on
small scales. 

In this paper, we emphasized that $\mu$ can be small for the
primordial suppression
scenario. However, ultimately it will be interesting to
study the small-scale problems by N-body simulations for a
variety of primordial spectra consistent with existing
constraints from, e.g., Lyman-$\alpha$ observation,
simultaneously calculating $\mu$ for each spectrum, possibly
taking into account baryonic processes.

\vspace{-4mm}
\begin{acknowledgments}
\vspace{-3mm}
\small{TN acknowledges useful discussions with Teruaki Suyama and
Jun'ichi Yokoyama.  TN was supported by
Grant-in-Aid for JSPS Fellow No.\ 25.8199 and JSPS 
Postdoctoral Fellowships for Research Abroad. JC is
supported by the Royal Society as a Royal Society University
Research Fellow at the University of Manchester, UK.  MK is
supported by the Simons Foundation, NSF Grant No.\ PHY-1214000,
and NASA ATP Grant No.\ NNX15AB18G.}
\end{acknowledgments}

\end{document}